\documentclass[aps,prb,twocolumn]{revtex4}
\usepackage{epsfig}

\newcommand{\Sp}{\mbox{{\bf S}}}
\newcommand{\Spg}{\mbox{{\bf S}}}
\newcommand{\be}{\begin{equation}}
\newcommand{\ee}{\end{equation}}
\newcommand{\ben}{\begin{eqnarray}}
\newcommand{\een}{\end{eqnarray}}
\newcommand{\ra}{\rangle}
\newcommand{\la}{\langle}
\newcommand{\im}{{\rm i}}
\newcommand{\Rg}{{\bf R}}
\newcommand{\kk}{{\bf k}}

\newcommand{\psbild}[1]{#1}  

\begin{document}

\title{Influence of quantum fluctuations on zero-temperature 
       phase transitions between collinear and noncollinear states 
       in frustrated spin systems}

\author{Sven E.~Kr\"uger} 
\author{Johannes Richter}
\affiliation{Institut f\"ur Theoretische Physik,
             Universit\"at Magdeburg,\\
             P.O.Box 4120, 39016 Magdeburg, Germany}

\date{\today}

\begin{abstract}
We study a square-lattice spin-half 
Heisenberg model where frustration is introduced by competing 
nearest-neighbor bonds of different signs. 
We discuss the influence of quantum fluctuations on the nature of the
zero-temperature phase transitions from phases with collinear magnetic 
order at small frustration  
to phases with noncollinear spiral order at large frustration.
We use the coupled cluster method (CCM) for high orders
of approximation (up to LSUB6) and the  exact diagonalization of
finite systems (up to 32 sites) to calculate ground-state properties.
The role of
quantum fluctuations is examined by comparing the ferromagnetic-spiral and the
antiferromagnetic-spiral transition  within the same model.
We find clear evidence that quantum fluctuations prefer collinear order and
that they may favour a first order transition instead of a second order
transition in case of no quantum fluctuations.
\end{abstract}

\maketitle

{\em Introduction.}
While quantum fluctuations do not influence the critical properties
of phase transitions at $T>0$ they play an important role at $T=0$ and
can yield to quantum phase transitions, which have attracted a lot 
of attention in recent times (see, e.g., Ref.~\onlinecite{qpt}).
Quantum fluctuations arise due to Heisenberg's uncertainty principle and
play a similar role as thermal fluctuations (for $T>0$) in classical 
transitions.
The spin-half Heisenberg model is a basic model which shows strong quantum
fluctuations in the antiferromagnetic case. While the ground-state of 
the pure Heisenberg antiferromagnet (HAFM) on the square lattice shows 
N\'{e}el long-range order\cite{manou91} (LRO), a competition of bonds 
can increase quantum fluctuations and
may result in rotationally invariant paramagnetic states, suppressing
the (collinear) N\'eel order.
This is demonstrated by recent experiments on (quasi-)two-dimensional
Heisenberg systems, like CaV$_4$O$_9$ (see, e.g., 
Refs.~\onlinecite{tanigu95,troyer96}) or  SrCu$_2$(BO$_3$)$_2$
(see, e.g., Refs.~\onlinecite{kageyama99,koga00a}).

Besides local singlet formation magnetic 
frustration is an important mechanism to drive zero-temperature 
transitions. In the classical Heisenberg 
model strong frustration often leads to 
noncollinear  (e.g., spiral) spin states which may or may not 
have counterparts in the quantum case. It is
generally argued that quantum fluctuations prefer a collinear ordering.
A typical example is the frustrated spin-half $J_1-J_2$ model
on the square lattice (see, e.g., 
Refs.~\onlinecite{ri93,oitmaa96,Bishop6,sushkov,sorella00}). 
Here the  classical version of the 
$J_1-J_2$ model has a continuously degenerate ground state for $J_2 > J_1/2$,
but quantum fluctuations can remove this degeneracy yielding to a collinear
state (``order from disorder'' phenomenon, 
see, e.g., Refs.~\onlinecite{prakash90,kubo90}).
Moreover quantum fluctuations can shift the critical point
of a collinear-noncollinear transition so that the collinear state can survive
into a region where classically it is already unstable.\cite{ivanov98,richter00,krueger00}

In this paper we extend  our previous work,\cite{krueger00}
where we have studied the transition from a collinear N\'{e}el order to
noncollinear spiral order in a frustrated spin-half HAFM
and consider now the transition form a collinear ferromagnetic order to a
noncollinear spiral order within the same model. 
While in the classical version of the model both situations can be mapped
onto each other the quantum model behaves basically different in both cases.
That is because of the different nature of the collinear state: While the 
quantum N\'{e}el state on two-dimensional lattices 
exhibits strong quantum fluctuations (the sublattice magnetization of the
HAFM on the square lattice is only about 60\% of its classical value)
the ferromagnetic state is the same for the quantum and the classical model
and there are no quantum fluctuations in this state.   

We use the coupled cluster method\cite{coester58,bishop91}
(CCM) and exact diagonalization of finite systems 
to calculate the ground state.
The CCM is a very powerful method and, particularly
high-order implementations of this method can be used to obtain 
a consistent description of various aspects of quantum spin systems
(for an overview see, for example, 
Refs.~\onlinecite{bishop91a,Bishop2,Farnell2,zeng98,bishop99,bishop00}).
We note that another important method for spin systems, 
the quantum Monte Carlo method, cannot be used for frustrated spin systems 
since it suffers from the minus sign problem.

\begin{figure}[ht]
  \psbild{\centerline{\epsfysize=7cm \epsfbox{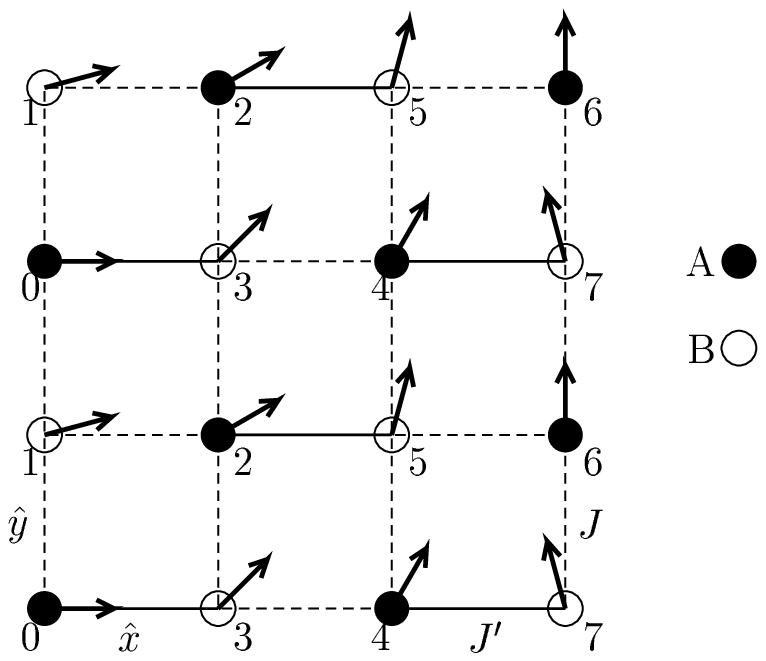}}}
  \caption{Illustration of the classical spiral state for the square-lattice
           Heisenberg model of Eq.~(\protect{\ref{ham}}),
           with two kinds of regularly distributed nearest-neighbour
           exchange bonds, $J$ (dashed lines) and $J'$ (solid lines). 
           The spin orientations at $A$ and $B$ lattice sites are defined
           by the angles $\theta_n=n\Phi$
           where $n=0, 1, 2, ...$, and $\Phi$ is the
           characteristic angle of the spiral state.
           The state is shown for
           $\Phi=\pi/12$ and $n=0, 1,\dots, 7$ and refers to the ferromagnetic
           case ($J<0$) with a $J'>|J|/3$. For the antiferromagnetic case ($J>0$
           and $J'<-J/3$) all spins on the B sublattice are reversed.}
  \label{fig1}
\end{figure}

{\em The model.} 
We consider a spin-half Heisenberg model on a square lattice 
with two kinds of nearest neighbour bonds $J$ and $J'$, as shown in 
Fig.~\ref{fig1},
\begin{equation}\label{ham}
        H  =  J\sum_{\langle ij\rangle_1}{\bf S}_i\cdot{\bf S}_j
+J'\sum_{\langle ij\rangle_2}{\bf S}_i\cdot {\bf S}_j
.\end{equation}
The sums over $<ij>_1$, and $<ij>_2$ represent sums over the nearest-neighbour
bonds shown in Fig.~\ref{fig1} by dashed and solid lines, respectively.
Each square-lattice pla\-quette consists of three $J$ bonds and one $J'$ bond.
A model with such a zigzag pattern has been treated with various 
methods.\cite{krueger00,singh88,ivanov96}

In this paper we consider only the cases in which $J$ and 
$J'$ have {\em different} signs
(i.e., one bond is ferromagnetic while the other is antiferromagnetic) so that
the plaquettes are frustrated. The case with antiferromagnetic $J$ bonds
(i.e., $J>0$ and hence $J'<0$) has been studied previously using linear spin 
wave theory,\cite{ivanov96} exact diagonalization and coupled cluster 
method.\cite{krueger00} We therefore focus in this paper our attention 
mainly on the ferromagnetic case (i.e., $J<0$ and $J'>0$) but compare 
the obtained results with those of the antiferromagnetic case.

{\em Classical ground state.} 
We  consider the ground state of the classical version of model
(\ref{ham}), 
i.e., the spins ${\bf S}_i$ are assumed to be classical vectors.  
For $|J'|<|J|/3$ (and $J$ and $J'$ having different signs)
the ground state of (\ref{ham}) is collinear
(i.e., ferromagnetic or antiferromagnetic depending on the sign of
$J$). At the critical point $J_c'=-J/3$,
a second-order transition takes place from the collinear state to a 
noncollinear state of spiral nature 
(see Fig.~\ref{fig1}), with a characteristic pitch 
angle $\Phi=\pm|\Phi_{\rm cl}|$ given by
\be\label{phi}
   |\Phi_{\rm cl}|=\left\{\begin{array}{ll}
                                                            0 & \quad |J'|<\frac{|J|}{3} \\
    \arccos\left(\frac{1}{2}\sqrt{1+\frac{1}{|J'|}}\; \right) & \quad |J'|\ge \frac{|J|}{3} \\
    \end{array}\right.
.\ee
Note that for $\Phi=0$ this is the collinear state.

The spins $\Sp_A$ and $\Sp_B$, belonging
to the $A$ and $B$ sublattices respectively, can be expressed in terms of
the spiral $\kk$ vector\cite{krueger00} with $\kk=(2\Phi,0)$ 
(and see Fig.~\ref{fig1}).
We note that this spiral state is incommensurate in the $x$-direction.
We also note that for the classical model the antiferromagnetic case can
be transformed into the ferromagnetic case 
by the simultaneous substitution $J \to -J$, $J' \to -J'$, 
${\bf S}_{i \in B} \to -{\bf S}_{i \in B}$. 
Hence the physics for both cases is classically the same.

{\em Calculation of the quantum ground state.}
To calculate the quantum ground state of the Hamiltonian (\ref{ham}) we
use the CCM. Details concerning the treatment of the model (\ref{ham}) 
with the CCM are given in Ref.~\onlinecite{krueger00}. 
We use the CCM for high orders of approximation up to LSUB6 
(using 1638 fundamental configurations).

We further exactly diagonalize finite lattices of rectangular shape 
($L_x \times L_y =4\times 4,\; 6\times 4,\;8\times 4$) using periodic boundary conditions.
The longer side $L_x$ of the rectangle corresponds to the direction of the 
$J'$ bonds and so
we can diminish the influence of the boundary conditions by an increase 
of $L_x$.

{\em The collinear-noncollinear transition.}
While classically we have always a second-order phase transition from 
collinear order to noncollinear order at $J'_c=-J/3$, we obtain for the quantum 
case a different behaviour for the ferromagnetic and the antiferromagnetic
case. 

\begin{figure}[ht]
  \psbild{\centerline{\epsfysize=7cm \epsfbox{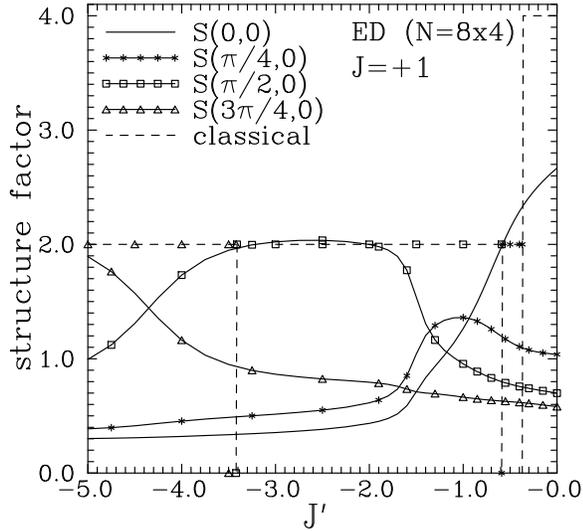}}}
  \caption{Ground-state structure factor 
           $S(\kk)\propto\sum_{i,j\in A}e^{\im(\Rg_j-\Rg_i)\cdot\kk}\la\Spg_i\cdot\Spg_j\ra$
           (i.e., the summation is taken over one sublattice)
           for a $8\times 4$ lattice 
           (with antiferromagnetic $J=+1$) for the quantum and the 
           classical case for various spiral vectors $\kk$.}
  \label{fig_str_af}
\end{figure}

\begin{figure}[ht]
  \psbild{\centerline{\epsfysize=7cm \epsfbox{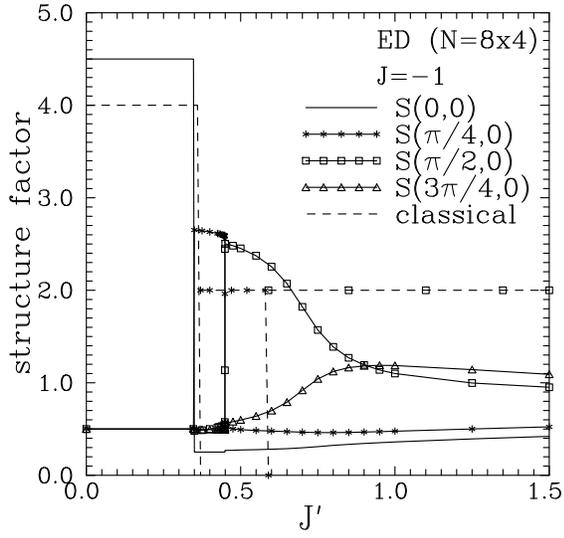}}}
  \caption{Ground-state structure factor $S(\kk)$ 
           (see Fig.~\protect{\ref{fig_str_af}})
           for a $8\times 4$ lattice 
           (with ferromagnetic $J=-1$) for the quantum and the classical case
           for various spiral vectors $\kk$.}
  \label{fig_str_fm}
\end{figure}

\begin{figure}[ht]
  \psbild{\centerline{\epsfysize=7cm \epsfbox{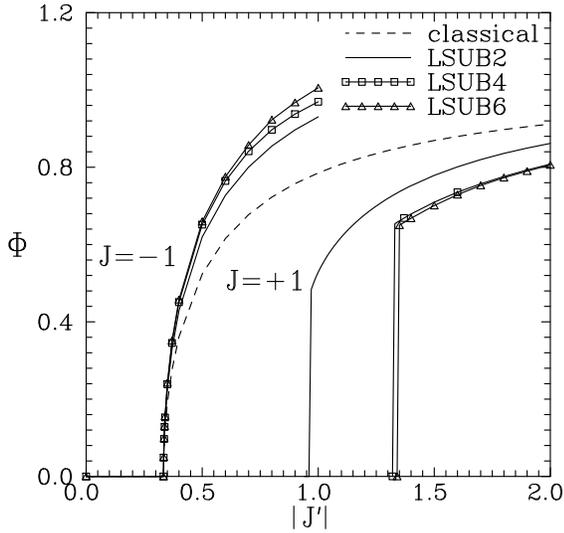}}}
  \caption{Pitch angle $\Phi$ versus $|J'|$ for the quantum and the classical
           case. While $\Phi$ is classically the same for the ferromagnetic 
           case ($J=-1$, $J'>0$) and
           for the antiferromagnetic case ($J=+1$, $J'<0$) 
           the quantum pitch angle is different for both cases. 
           The curves left of the classical (dashed) curve belong to $J=-1$ 
           and those right of it to $J=+1$. 
           Note that for the ferromagnetic case
           ($J=-1$) for $J'>1$ the pitch angle $\Phi$ becomes 
           meaningless, since the spiral
           order is already destroyed in this region 
           (c.f., Fig.~\protect{\ref{fig_m}}).}
  \label{fig_phi}
\end{figure}

Using the CCM we find for the antiferromagnetic case ($J=+1$) indications 
for a shift of this critical point to a value $J'_c\approx -1.35$
(see Fig.~\ref{fig_phi}). On the other hand for the ferromagnetic case
 ($J=-1$) we do not find such a shift (see Fig.~\ref{fig_phi}).
The exact diagonalization (ED) data of the structure factor $S(\kk)$
(see Figs.~\ref{fig_str_af} 
and \ref{fig_str_fm}) agree to these findings. For $J=+1$ 
(see Fig.~\ref{fig_str_af}) the collinear N\'eel order [$\kk=(0,0)$] becomes
unstable against the noncollinear spiral order [$\kk=(\pi/4,0)$]
in the classical model for
$J'\lesssim -0.36$ but in the quantum case only for $J'\lesssim -0.95$. 
The situation for the ferromagnetic case ($J=-1$) is again different. 
Here the results of the structure factor (see Fig.~\ref{fig_str_fm}) show that 
the transition from $\kk=(0,0)$ (collinear ferromagnetic order) to $\kk=(\pi/4,0)$ 
(spiral order) takes place at nearly the same value of $J'\approx 0.36$ 
for both, the classical case and the quantum case. 

\begin{figure}[ht]
  \psbild{\centerline{\epsfysize=7cm \epsfbox{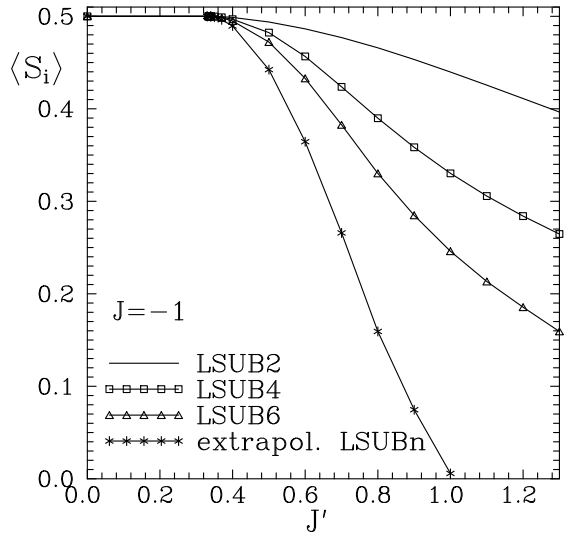}}}
  \caption{On-site magnetic moment versus $J'$ for the ferromagnetic case 
           ($J=-1$) calculated within the  CCM-LSUB$n$ approximations and 
           extrapolated to $n=\infty$ (The extrapolation is done
           as described in Ref.~\cite{krueger00}).} 
  \label{fig_m}
\end{figure}

Taking the deviation of the on-site magnetic moment $\la S_i\ra$
from its classical value $\la S_i\ra_{\rm cl}=1/2$ as an indication for
the degree of quantum fluctuations we can compare the strength of 
quantum fluctuation near the collinear-noncollinear transitions for both,
the antiferromagnetic and the ferromagnetic case. As reported in 
Ref.~\onlinecite{krueger00} for $J=+1$ the quantum fluctuations are particularly 
strong near the antiferromagnetic-spiral transition leading to an 
on-site magnetic moment less then 20\% of its classical value.
On the other hand, it can be seen  from Fig.~\ref{fig_m} that the 
on-site magnetic moment takes its classical value $1/2$ up to 
$J'\approx 0.36$ for $J=-1$ and therefore virtually {\em no} quantum 
fluctuations occur at the ferromagnetic-spiral transition.
Hence the shift of the critical
$J_c'$ in the antiferromagnetic case can clearly 
be attributed to the strong quantum fluctuations. 
In general our findings are consistent with
the statement that quantum fluctuations (which we have in the 
antiferromagnetic case only) prefer a collinear ordering, so that in this
case the quantum collinear state can survive  
into a frustrated region where classically the collinear state 
is already unstable.

We further note an agreement between the CCM results and the ED results
beyond the critical $J'_c$. By examining the structure factors
(see Figs.~\ref{fig_str_af} and \ref{fig_str_fm}) we find that 
for the antiferromagnetic (respectively ferromagnetic) case
the transitions to a spiral state with a greater $\kk$ vector (i.e.,
with a greater pitch angle $\Phi$) occur in the quantum model 
always at a greater
(respectively smaller) {\em absolute} value of $J'$ then the corresponding 
classical transitions. This agrees with the CCM results of the pitch angle
(see Fig.~\ref{fig_phi}), 
where we have $\Phi_{\rm qm} < \Phi_{\rm cl}$ (respectively $\Phi_{\rm qm} >
\Phi_{\rm cl}$). 

The discussion given above corresponds to our finding concerning the order
of the transition. 
Clearly in the ferromagnetic case ($J'=-1$) both the classical and the
quantum model show a second-order phase transition 
(see  Figs.~\ref{fig_phi} and \ref{fig_m}). 
On the other hand it was discussed in Ref.~\onlinecite{krueger00}
that the collinear-noncollinear transition
in the antiferromagnetic case ($J=+1$) is probably 
a first-order phase transition for the quantum model (c.f., Fig.~\ref{fig_phi}) 
in difference to the second-order transition in the classical case.

{\em Formation of local singlets.}
For sufficient strong antiferromagnetic $J'$ bonds the system
(\ref{ham}) is characterized by a tendency to singlet pairing of the two
spins coupled by a $J'$ bond and hence the long-range magnetic
(collinear or noncollinear) order is destroyed. 
We obtain clear indications of a second-order phase transition to
a quantum paramagnetic dimerized phase at 
a certain critical value of $J'=J_s'$. 
For $J'>J'_s$ the on-site magnetic moment $\la S_i\ra$ becomes
zero. For the ferromagnetic case ($J=-1$) 
we find $J'_s\approx 1$ using the extrapolated CCM-LSUB$n$ results
(see Fig.~\ref{fig_m}). We may also consider the inflection points of 
$\la S_i\ra$ versus $J'$ for the LSUB$n$ approximations,
assuming that the true curve will have a negative curvature up to the
critical point. We find the corresponding inflection points at $J'\approx 1.2$ 
($n=2$), $J'\approx 0.76$ ($n=4$) and $J'\approx 0.74$ (n=6), indicating 
a critical $J'_s$ even slightly smaller then $J'_s\approx 1$.

\begin{figure}[ht]
  \psbild{\centerline{\epsfysize=7cm \epsfbox{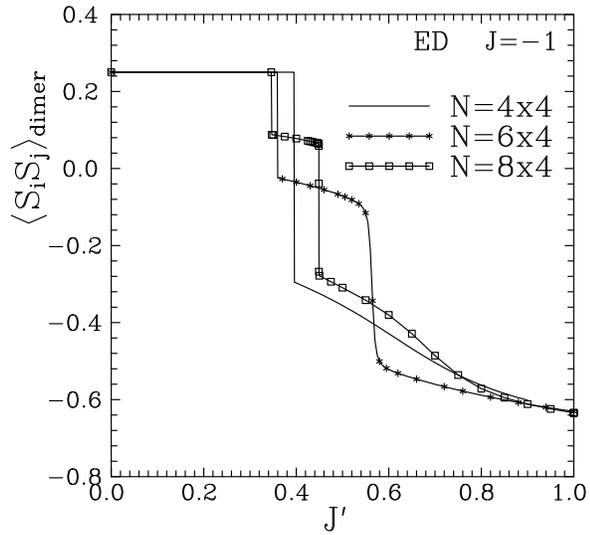}}}
  \caption{Nearest-neighbour spin-spin correlation of the two spins connected 
           via a $J'$ bond versus $J'$ for the ferromagnetic case
           using exact diagonalization (ED) data.}
  \label{fig_sij}
\end{figure}

The ED data give a similar approximation of $J'_s$:
For $J'\approx 1$ finite-size effects in 
spin-spin correlations disappear almost completely 
(see for illustration Fig.~\ref{fig_sij}). This indicates that spin-spin
correlations are short ranged with a length scale smaller than the
size $L_x$. 

We note that for the antiferromagnetic case ($J=+1$) the
strength of antiferromagnetic $J'$ needed for breaking N\'{e}el order by
formation of local singlets is much larger
($J_s'\approx 3$, see Ref.~\onlinecite{krueger00}). The lower critical 
$J_s'\approx 1$ in the ferromagnetic case is due to 
frustration which assists local singlet formation 
(c.f., Ref.~\onlinecite{gros95,wei97}).

{\em Summary.}
Using the CCM and the ED technique we have studied the influence of quantum
fluctuations on zero-temperature transitions between collinear 
ordered and noncollinear ordered states in a frustrated spin-half 
square-lattice Heisenberg model with two kinds of 
nearest-neighbour exchange bonds. 
The frustration drives a second-order transition between collinear
(antiferro- or ferromagnetic) and noncollinear (spiral) states
in the classical model.
For the quantum model the CCM provides a consistent description of collinear,
noncollinear, and disordered phases, while some other 
standard techniques (e.g., QMC) are not applicable. 
We find a strong influence of quantum
fluctuations on the nature of the collinear-noncollinear transition, and
quantum fluctuations (which favour collinear ordering) may change the
second-order classical transition to a first-order quantum transition. If
quantum fluctuations are suppressed in the collinear phase of the quantum
model, the transition to the spiral phase is similar for the quantum and
for the classical model.

{\em Acknowledgments.}
We thank the Deutsche Forschungsgemeinschaft 
(Ri 615/9-1) for its support. We are indebted to J.~Schulenburg for
numerical assistance.


\end{document}